\documentclass[nofootinbib,amsmath,amssymb,aps,prd,showpacs]{revtex4}
\usepackage[colorlinks=true, pdfstartview=FitV, linkcolor=blue, citecolor=red, urlcolor=magenta]{hyperref}
\usepackage{graphicx,color}
\usepackage{latexsym}
\usepackage{amsmath}
\usepackage{amsfonts}
\usepackage{amssymb}

\newcommand{\be}{\begin{equation}}
\newcommand{\ee}{\end{equation}}
\newcommand{\bea}{\begin{eqnarray}}
\newcommand{\eea}{\end{eqnarray}}

\newcommand{\pa}{\partial}

\newcommand{\bb}{\bibitem}
\def\pls{\partial\!\!\!/}
\def\bb{\bibitem}

\def\ps{p\!\!\!/}

\def\ns{n\!\!\!/}

\def\ds{\partial\!\!\!/}

\def\bb{\bibitem}
\newcommand{\dslash}{\partial\!\!\!/}
\newcommand{\aslash}{A\!\!\!/}

\newcommand{\ben}{\begin{eqnarray}}
\newcommand{\een}{\end{eqnarray}}

\begin{document}
\title{The 4D-2D projection of Lorentz-violating Myers-Pospelov QED}


\author{$^{a}$F. A. Brito, $^{b}$M. S. Guimaraes, $^{a}$E. Passos,
$^{a}$P. Sampaio, $^{c}$C. Wotzasek} \affiliation{$^{a}$Departamento de F\'\i sica,
Universidade Federal de Campina Grande, Caixa Postal 10071,
58109-970  Campina Grande, Para\'\i ba,
Brazil\\
$^{b}$Departamento de F\'\i sica Te\'orica, Instituto de F\'\i sica, Universidade do Estado do Rio de Janeiro, Rua S\~ ao Francisco Xavier 524, 20550-013 Maracan\~ a, Rio de Janeiro, Brazil\\
$^{c}$Instituto de F\'\i sica, Universidade Federal do Rio de Janeiro,
\\Caixa Postal 21945, Rio de Janeiro, Brazil
}
\email{fabrito@df.ufcg.edu.br,  msguimaraes@uerj.br, passos@df.ufcg.edu.br, clovis@if.ufrj.br}

\begin{abstract}
We consider the four-dimensional quantum electrodynamics extended with Myers-Pospelov Lorentz-violating dimension-5 operators to investigate 4D-2D projection. In projecting out the 4D theory down to a 2D theory we get analogs of these operators. Namely, we obtain a new two-dimensional theory with corresponding scalar and fermionic 2D Myers-Pospelov Lorentz-violating dimension-3 operators. New defect structures can also be found from this new projected out 2D scalar sector. Furthermore, we also show that this 2D scalar sector can be also radiatively induced through the new 2D fermionic sector. 
\end{abstract}
\maketitle
\newpage

\section{Introduction}

In order to find a new physics, important features on Lorentz symmetry violation with high derivative or high dimension operators have been recently explored in quantum gravity \cite{horava} and field theory \cite{MP_01,kost2012,mariz,sar,aabp,mariz2,Scarpelli:2012kt}. As we shall see later, these operators develop similar characteristics of Lorentz-violating dimension-4  operators initiated in the seminal papers \cite{kost}. However, we shall focus on low-dimensional effects from the four-dimensional theories with high derivative Lorentz-violating operators such as the four-dimensional quantum electrodynamics extended with Lorentz-violating dimension-5 operators proposed by Myers and Pospelov \cite{MP_01}. In this work we initially consider the gauge sector of this modified theory in order to derive the physics in two dimensions with analogs of these Lorentz-violating operators. Namely, we are able to obtain scalar and fermionic 2D Myers-Pospelov dimension-3 operators. Another purpose of this work is to study whether the same scalar 2D  Myers-Pospelov term can be also radiatively induced into an effective action from the resulting fermionic 2D Myers-Pospelov sector obtained through the 4D fermionic sector projected into two dimensions. 

In two dimensions, we have that the Lorentz symmetry violation can be attributed to the scalar sector
added to an effective theory that also can characterize a preferred direction for two-dimensional space-time. Models of this nature was suggested and studied in the context of the defect structures \cite{dio, Bazeia:2010yp, Dutra:2010uk}, deformation of the canonical structure and {UV/IR} duality \cite{clo} and radiative induction via one-loop polarization tensor in two dimensions that is independent of regularization scheme \cite{Passos:2008qh}. 

In the present study the scalar Lorentz-violating terms in two dimensions are obtained both via radiative corrections and four-dimensional projection down to two dimensions. The first case produces a scalar 2D Myers-Pospelov term induced through fermionic 2D Myers-Pospelov term.
Also, it is naturally expected, that low dimensional Lorentz-symmetry violating effective actions naturally arises in the process of dimensional reduction of the four-dimensional  theories  which was earlier shown in \cite{man, man1,CS3,Casana:2011vh} to give the 3D Lorentz-violating action.

The extended QED in four-dimension is given as
\bea\label{qed_01} 
{\cal L}_{qed_{4D}}=-\frac14 F_{\mu\nu}F^{\mu\nu} + g\varepsilon^{\mu\nu\rho\sigma}n_{\mu}(n\cdot \pa)^{2}\partial_{\nu}A_{\rho}A_{\sigma}+
 \bar{\psi}\left(i\dslash  +g^\prime(n\cdot \pa)^{2}\ns \Gamma_5 - M\right)\psi - e\bar\psi \aslash\psi,
\eea
where the $n_{\mu}$ is the dimensionless constant four-vector, $e$ is the minimal coupling constant, $g  = \eta/M_P $ and $g^{\prime}= \eta^{\prime}/M_P$ with $\eta, \eta^\prime$  being dimensionless parameters.
These dimension-5 operators in the electromagnetic and fermionic sectors are Myers-Pospelov-type operators. Both operators break CPT symmetry being $\eta$ and $\eta^{\prime}$ charge conjugation even, 
and $M_P$ is the mass where new physics such as Lorentz and CPT symmetries violation emerges. 

We shall firstly discuss the emergence of a 2D theory via the gauge field sector of above Lagrangian from 4D-2D dimensional projection. In this point, we shall focus on low-dimensional effects from the four-dimensional theories with high derivative Lorentz-violating operators. We take advantage of this procedure to obtain a 2D theory with two scalar fields coupled to dimensional-3 operators. This aims at to find new topological solutions. It turns that the equations of motion are the same as those ones previously found in other Lorentz-violating scalar field scenarios \cite{dio}. This is because, the projected out high derivative 2D dimensional theory reduces to a lower derivative theory. Secondly we shall focus on the study whether the same emerging 2D effective action can be induced as an effective action from the resulting 2D fermion field sector obtained through the 4D fermionic sector of above Lagrangian projected into two dimensions. 


This paper is organized as follows. In section \ref{s1}, we develop the dimensional projection of a 4D theory into a 2D one for the gauge sector of the extended QED. In section \ref{s2}, we take advantage of this procedure to obtain a 2D theory with two scalar fields coupled to dimensional-3 operators. We point out the possibility of finding new defect structures in the projected out 2D Myers-Pospelov scalar model.  In section \ref{s3}, we shall address the issue of 4D-2D projection in the fermion field sector. In section \ref{s4}, we revisit the problem on induced 2D effective theory via radiative corrections at one loop contribution via derivative expansion method.
Our starting point here is the 2D fermion sector projected out from the 4D theory. Finally, in section \ref{s5} we present our conclusions.

\section{4D-2D projection: Gauge Field Sector}\label{s1}
In this section, we now focus on dimensional projection out of the four-dimensional gauge field components of a 4D theory into a 2D theory. 
Firstly, the Maxwell term assumes the form
\begin{align}
-\frac14 F_{\mu\nu}F^{\mu\nu} &= -\frac14 \left( F_{ij}F^{ij} + 2F_{0a}F^{0a} + 2F_{3a}F^{3a} + F_{ab}F^{ab}\right)\nonumber\\
&= -\frac14 \left( F_{ij}F^{ij} + 2\partial_0 A_a \partial^0 A^a + 2\partial_3 A_a \partial^3 A^a \right)\;, \label{projMax}
\end{align}
where $i,j = 0,3$, $a,b = 1,2$, and the projection implies to say that $\partial_1 = \partial_2 = 0$ when acting on any field (therefore $F_{ab}=0$). We will work with the metric signature $(+,-,-,-)$, so that $A^{a} = -A_{a}$. Now renaming $A_1 = \phi$ and $A_2 = \chi$ we end up with
\begin{align}
-\frac14 F_{\mu\nu}F^{\mu\nu} = -\frac14 F_{ij}F^{ij} + \frac12 \partial_i\phi\partial^i\phi + \frac12 \partial_i\chi\partial^i\chi\;. \label{projMax2}
\end{align}
This is a Maxwell theory plus two massless scalar field theory in 2D.

Now let us discuss the Myers - Pospelov term. Under the same conditions described above we obtain:
\begin{align}
g\varepsilon^{\mu\nu\rho\sigma}n_{\mu}(n\cdot \pa)^{2}\partial_{\nu}A_{\rho}A_{\sigma} &= g\big(n_{i}\pa^{i}+n_{a}\pa^{a}\big)^{2}\big(\varepsilon^{ij\rho\sigma}n_{i}\partial_{j}A_{\rho}A_{\sigma} + \varepsilon^{ai\rho\sigma}n_{a}\partial_{i}A_{\rho}A_{\sigma}\big)\nonumber\\
&= g\big(n_{i}\pa^{i}+n_{a}\pa^{a}\big)^{2}\big(\varepsilon^{ij}n_{i}\chi\partial_{j}\phi   +\varepsilon^{ai\rho\sigma}n_{a}\partial_{i}A_{\rho}A_{\sigma}\big)
\;, \label{projCS}
\end{align}
where $\varepsilon^{ij}\equiv\varepsilon^{ij12}$. { Note that the first term is the one of  major interest to us. The second term will necessarily mix the 2D Maxwell's and scalar fields, but here for the sake of simplicity we shall not consider it. For that reason let us consider from now on the $n_a =0$ case.

The projection of the 4D gauge field sector down to a 2D theory is finally given by}
\begin{align}
{\cal L}_{2D} = -\frac14 F_{ij}F^{ij} + \frac12 \partial_i\phi\partial^i\phi + \frac12 \partial_i\chi\partial^i\chi\ + g\varepsilon^{ij}n_{i}(n_{j}\pa^{j})^{2}\chi\partial_{j}\phi \;. \label{projCFJ}
\end{align}
This resulting theory are made out of two-dimensional fields living in two dimensions. The new term we shall refer as to 2D Myers-Pospelov scalar model.

\section{Defect Structures in projected out 2D Myers-Pospelov scalar model}\label{s2}

By following the previous projected Lagrangian in two dimensions, in this section we propose a new theory with two real scalar fields with a higher derivative dynamics written in the form
\bea\label{v1}
{\cal L}_{2D}= \frac12 \partial_{\mu}\phi\partial^{\mu}\phi + \frac12 \partial_{\mu}\chi\partial^{\mu}\chi\ + g\varepsilon^{\alpha\beta}n_{\alpha}(n\cdot\pa)^{2}\chi\partial_{\beta}\phi- V(\phi,\chi),
\eea
where $\mu=0,1$ and $n_\mu=(n_0,n_1)$.
This effective theory can generate new effects on (non-)topological defects generated by two real scalar fields with a proper choice of the scalar potential $V(\phi,\chi)$.

Let us obtain the equations of motion. The equation of motion for the scalar fields $\phi$ and $\chi$ are, respectively, given by
\ben
\square\phi + g\epsilon^{\alpha\beta}n_\alpha(n\cdot\pa)^{2}\partial_\beta\chi + \frac{\partial V}{\partial\phi}=0,\nonumber\\
\square\chi - g\epsilon^{\alpha\beta}n_\alpha(n\cdot\pa)^{2}\partial_\beta\phi + \frac{\partial V}{\partial\chi}=0.
\een
We shall look now for static solutions. Now assuming $\phi=\phi(x)$ and $\chi=\chi(x)$ we obtain
\ben
-{\phi}^{\prime\prime}+gn_1^3\chi^{\prime\prime\prime}+\frac{\partial V}{\partial\phi}=0,\nonumber\\
-{\chi}^{\prime\prime}-gn_1^3\phi^{\prime\prime\prime}+\frac{\partial V}{\partial\chi}=0.
\een
The equations do not depend on time-like component $n_0$. Thus, if a priori one considers only time-like components there is no Lorentz violation effects on the solutions in the static regime. 
We also note that for a constant (or in the absence of) potential $V$, the above equations, up to constants that can be eliminated via boundary conditions, are simply reduced to
\ben\label{eqs-new}
{\phi}-gn_1^3\chi^{\prime}=0,\nonumber\\
{\chi}+gn_1^3\phi^{\prime}=0.
\een
This equations are equivalent to those found in Ref.~\cite{dio} in the same regime. This shows that the Myers-Pospelov operator in the static regime with no potential contribution in two dimensions becomes a reducible operator.  They can be found from the use of the following two-dimensional  Lagrangian \cite{Passos:2008qh}
\ben
{\cal L}=\frac12 \partial_{\mu}\phi\partial^{\mu}\phi + \frac12 \partial_{\mu}\chi\partial^{\mu}\chi+\kappa_\alpha\epsilon^{\alpha\beta}\chi\partial_\beta\phi,
\een
as long as the time-like component $n_0$ is turned on.

Now returning to equations (\ref{eqs-new}) we find that they are satisfied by oscillatory solutions 
\ben
\phi=\cos{kx},\qquad \chi=\sin{kx}, \qquad k=\frac{1}{gn_1^3}.
\een
In the following we shall consider non-trivial potentials $V$. If we assume $V$ satisfies the following relations
\ben
\int{\frac{\partial V}{\partial \phi}dx}=-\frac{\partial U}{\partial \chi},\qquad \int{\frac{\partial V}{\partial \chi}dx}=\frac{\partial U}{\partial \phi},
\een
or
\ben
{\frac{\partial \cal V}{\partial \phi}}=-\frac{\partial U}{\partial \chi},\qquad {\frac{\partial \cal V}{\partial \chi}}=\frac{\partial U}{\partial \phi}.
\een
Thus, we can reduce the equations of motion to the following 
\ben
-{\phi}^{\prime}+gn_1^3\chi^{\prime\prime}-\frac{\partial U}{\partial\chi}=c_1,\nonumber\\
-{\chi}^{\prime}-gn_1^3\phi^{\prime\prime}+\frac{\partial U}{\partial\phi}=c_2,
\een
where $c_1, c_2$ are integration constants that we shall set to zero here.   
Let us recast these equations in the familiar form \cite{dio}
\ben
-\phi^{\prime\prime}-k{\chi}^{\prime}+\frac{\partial \tilde{U}}{\partial\phi}=0,\nonumber\\
-\chi^{\prime\prime}+k{\phi}^{\prime}+\frac{\partial \tilde{U}}{\partial\chi}=0,
\een
where $\tilde{U}=kU$. Now every solution found in \cite{dio} can be related to solutions of our system.

\section{4D-2D Projection: fermion field sector}\label{s3}

 Let  us now investigate the dimensional projection in the fermion sector of  the quantum electrodynamics extended with dimension-5 operators (\ref{qed_01}) whose fermionic Lagrangian is given by
\begin{align}
{\cal L}_{fermi_{4D}} = \bar{\psi}\left(i\dslash +g^\prime (n\cdot \pa)^{2}\ns \Gamma_5 - M \right)\psi - e\bar\psi\aslash\psi\;. \label{fermi4D}
\end{align}
We consider the following representation of the gamma matrices in $4D$:
\begin{align}
\label{gamma4D}
\Gamma_5 = \left( \begin{array}{cc}
-1 & 0 \\
0 & 1 \end{array} \right); \;\;\; \Gamma^{\mu} = \left( \begin{array}{cc}
0 & \sigma^{\mu}\\
\bar{\sigma}^{\mu} & 0 \end{array} \right);\;\; \sigma^{\mu} =(1, \overrightarrow{\sigma} );\;\;\; \bar{\sigma}^{\mu} =(1, -\overrightarrow{\sigma} )\;.
\end{align}
Now defining
\begin{align}
\label{psi}
\psi = \left( \begin{array}{c}
\psi_L \\
\psi_R \end{array} \right)\;,
\end{align}
and recalling that $\bar{\psi} = \psi^{\dagger}\Gamma^0$, we can write the Lagrangian (\ref{fermi4D}) as
\begin{align}
\label{fermi4D-matrix}
{\cal L}_{fermi_{4D}} = \left( \psi_L^{\dagger} \;\;  \psi_R^{\dagger} \right) \left( \begin{array}{cc}
i\bar{\sigma}^{\mu}\partial_{\mu} - e\bar{\sigma}^{\mu}A_{\mu} +g^\prime (n\cdot \pa)^{2} \bar{\sigma}^{\mu}n_{\mu}  & - M \\
- M & i\sigma^{\mu}\partial_{\mu} - e\sigma^{\mu}A_{\mu} - g^\prime (n\cdot \pa)^{2}\sigma^{\mu}n_{\mu} \end{array} \right) \left( \begin{array}{c}
\psi_L \\
\psi_R \end{array} \right)\;.
\end{align}
 As is well known, the sectors $\psi_L$ and $\psi_R$ decouple only if $M=0$. Setting $M=0$ and looking at, say, the sector $\psi_R$, we have
\begin{align}
\label{fermi4D-R}
{\cal L}^{R}_{fermi_{4D}} =  \psi_R^{\dagger} \left(i\sigma^{\mu}\partial_{\mu} - e\sigma^{\mu}A_{\mu} +g^\prime (n\cdot \pa)^{2} \sigma^{\mu}n_{\mu}\right)
\psi_R \;.
\end{align}
Now we proceed with the dimensional projection to the plane $x_0-x_3$ with this Lagrangian as the starting point. The gamma matrices in 2D can be defined as:
\begin{align}
\label{gamma2D}
\gamma_5 = \sigma^3 = \left( \begin{array}{cc}
1 & 0 \\
0 & -1 \end{array} \right); \;\;\; \gamma^{0} = \sigma^1 = \left( \begin{array}{cc}
0 & 1 \\
1 & 0 \end{array} \right); \;\;\; \gamma^{1} = -i\sigma^2 = \left( \begin{array}{cc}
0 & -1 \\
1 & 0 \end{array} \right)\;.
\end{align}
Notice that $\gamma^{1}$ is just the matrix associated with the spatial direction in two dimensions, as we chose this direction as $x_3$, it is convenient to rename the matrix as $\gamma^{3}$. This is just a change of labels. It is done so that the sums over $i=0,3$ make sense. The reason for this choice of gamma representation can be seen when we impose the conditions $\partial_1 = \partial_2 = 0$, we obtain
\begin{align}
\label{gammapartial}
\sigma^{\mu}\partial_{\mu} &= \sigma^{0}\partial_{0} +  \sigma^{3}\partial_{3} = \gamma^{0}\gamma^{0}\partial_{0} +  \gamma^{0}\gamma^{3}\partial_{3},\nonumber\\
n\cdot \pa&= n^{0}\pa_{0}+n^{3}\pa_{3}
\end{align}
which has the necessary structure to build the kinetic Dirac term in 2D. 

Following the same route as for the gauge field sector, we define $A_1 = \phi$ e $A_2 = \chi$ and $n_{a}=0$. We conclude that the Lagrangian (\ref{fermi4D-R}) when projected is given by:
\begin{align}
\label{fermi4D-R-proj}
{\cal L}^{R}_{ferm_{2D}} =  \bar{\psi}_R \left(i\gamma^{i}\partial_{i} - e\gamma^{i}A_{i} - e\phi -ie\gamma_5\chi +g^\prime (n^{i}\pa_{i})^{2} \gamma^{i}n_{i} 
\right)
\psi_R \;.
\end{align}
However, in order to study radiative corrections in the next section, we shall consider the $massive$ version of the massless  2D fermionic sector, given by
\bea\label{qed_02-1}
{\cal L}_{{2D}}=
\bar{\psi}( i \pls - m + g^{\prime}(n\cdot \pa)^{2}\ns)\psi- \tilde{e} \bar\psi (\phi+i\gamma_{5}\chi)\psi.
\eea
In two dimensions, $\tilde{e}$ has dimension of mass and we define $\tilde{e}^2={e^2}/({4\pi R^2})$ being $R$ 
related to the mass as $m\sim1/R$ and $e$ is the extended QED$_4$ coupling. The mass $m$ can be set to zero after making radiative corrections since the induced term has no explicit dependence with such a mass. See below. 
\section{Induced Scalar Effective Theory in 2D}\label{s4}

Let us explicitly study  the question of inducing the scalar 2D Myers - Pospelov  term into an effective action induced via
radiative correction. This issue
we will study in the following by using the derivative expansion of fermion determinants.

In this sense, we consider the fermion sector associated
to the Lagrangian (\ref{qed_02-1})  and integrate out the fermion fields into the functional integral to obtain the effective action for the scalar fields given by the following functional trace
\bea
\label{eq.003}
S_{eff}[n,\phi,\chi]=-i\,{\rm Tr}\,\ln\big[\ps - m -g^{\prime}(n\cdot p)^{2}\ns - \tilde{e}\big(\phi+ i \gamma_{5}\chi\big) \big]. 
\eea 
Applying the  method of derivative expansion, 
the functional trace in (\ref{eq.003}) can be properly manipulated in view of writing the effective action in the 
following form:
\bea
S_{eff}[n,\phi,\chi]=S_{eff}[n]+S_{eff}^{\,\prime}[n,\phi,\chi],
\eea 
where the first term is $S_{eff}[n]=-i\,{\rm Tr}\ln\big[\ps - m - g^{\prime}(n\cdot p)^{2}\ns\big]$,
which does not depend on the scalar fields. The  nontrivial
dynamics is only governed by the second term $S_{eff}^{\,\prime}[n,\phi,\chi]$,
which is given by the following power series 
\bea
\label{eq.004}
S_{eff}^{\,\prime}[n,\phi,\chi]=i\,{\rm Tr} \sum_{l=1}^{\infty}\frac1l
\Biggl[\frac{\tilde{e}}{\ps - m - g^{\prime}(n\cdot p)^{2}\ns}\,\big(\phi+ i \gamma_{5}\chi\big)\Biggr]^l.
\eea 
{ To induce the scalar 2D Myers - Pospelov term of the effective action} in (\ref{v1}) we should expand the equation
(\ref{eq.004}) up to second order in the scalar fields in the presence of the parameter $n_{\mu}$ that
controls the Lorentz symmetry violation. {Thus, we can now write the effective action in the form }
\bea
\label{eq.006}
S_{eff}[n,\phi,\chi]=\frac{\tilde{e}^{2}}{2}{\rm
Tr}\bigl[G_F(p)\phi(x) G_F(p)\gamma_{5}\chi(x)+G_F(p)\gamma_{5}\chi(x) G_F(p)\phi(x)\bigl], 
\eea
where $G_F(p)$ is the $n_{\mu}$-dependent propagator of the theory
defined as:
\bea
G_F(p)=\frac{i}{\ps- m - g^{\prime}(n\cdot p)^{2}\ns}.
\eea
Now, we expand the exact fermion propagator
$G_{F}(p)$ and take the leading order in $g^{\prime}$ as in the form
\bea
G_{F}(p)= S_{F}(p)+S_{F}(p)\big(-i g(n\cdot p)^{2}\ns\big)S_{F}(p)+\cdot\cdot\cdot\,\,\,{\rm with}\,\,\,S_{F}(p)=\frac{i}{\ps-m}.
\eea
Using the cyclic properties of the trace and the above result into the equation (\ref{eq.006}) we have
\bea
\label{eq.008}
S_{eff}[n,\phi,\chi]=-i g^{\prime}\tilde{e}^{2}{\rm
Tr}\bigl[S_F(p)(n\cdot p)^{2}\ns S_F(p)\gamma_{5}\chi(x)S_F(p)\phi(x)+S_F(p)(n\cdot p)^{2}\ns S_F(p)\phi(x)S_F(p)\gamma_{5}\chi(x)\bigl].
\eea
Now applying the main property of the derivative expansion method \cite{de} --- see also \cite{bpp, Brito:2008ec, Brito:2011hc}, we observe that any function of momentum can be
converted into a coordinate dependent quantity as follows 
\bea
\label{eq.081}
\phi(x)f(p)=\big(f(p-i\pa)\phi(x)\big).
\eea
The parenthesis on the right hand side merely emphasizes that the derivatives act only on $\phi(x)$.
In this case, the { two-point contribution to the effective action becomes}
\bea
\label{eq.00881}
S_{eff}[n,\phi,\chi]=-i g^{\prime}\tilde{e}^{2}{\rm
Tr}\big[(n\cdot p)^{2}\!\bigl[S_F(p)\ns S_F(p)\gamma_{5}\big(S_F(p-i\pa)\chi(x)\big)\phi(x)+S_F(p)\ns S_F(p)\big(S_F(p-i\pa)\phi(x)\big)\gamma_{5}\chi(x)\bigl]\big],
\eea
where
\bea\label{10}
&&S_{F}(p-i\pa)\chi(x)=\frac{i}{\ps-i\ds-m}\chi(x)=S_F(p)\chi(x)+S_{F}(p)\ds S_{F}(p)\chi(x)+\nonumber\\&&S_{F}(p)\ds S_{F}(p)\ds S_{F}(p)\chi(x)+S_{F}(p)\ds S_{F}(p)\ds S_{F}(p)\ds S_{F}(p)\chi(x)+\cdot\cdot\cdot.
\eea
Now, we can write the equation (\ref{eq.00881}) up to third order in the derivatives and in the lowest-order of  the coupling constant $g$, in the form
\bea\label{i1}
&&S_{eff}[n,\phi,\chi]=-ig^{\prime}\tilde{e}^{2}\int d^{2}x\int\frac{d^{2}p}{(2\pi)^{2}}\frac{(n\cdot p)^{2}}{(p^{2}-m^{2})}\times\nonumber\\&&
\Big[\big({\rm tr}\big[\ds S_{F}(p)\ds S_{F}(p)\ds S_{F}(p)S_{F}(p)\ns\gamma_{5}\big]\chi(x)\big)\phi(x)+
\big({\rm tr}\big[\ns S_{F}(p)S_{F}(p)\ds S_{F}(p)\ds S_{F}(p)\ds \gamma_{5}\big]\phi(x)\big)\chi(x)\Big]
\eea
where the symbol tr denotes the trace of the product of the gamma matrices. { As one knows from the gamma matrices algebra} and relation for trace ${\rm tr}\big(\gamma^{\mu}\gamma^{\nu}\gamma_{5}\big)=2\varepsilon^{\mu\nu}$, we can  write
the expression (\ref{i1}) as
\bea\label{i2}
&&S_{eff}[n,\phi,\chi]=-2ig^{\prime}\tilde{e}^{2}\int d^{2}x \,n_{\alpha}\big(\pa_{\beta}\pa_{\mu}\pa_{\nu}\chi(x)\big)\phi(x)\int\frac{d^{2}p}{(2\pi)^{2}}\frac{(n\cdot p)^{2}}{(p^{2}-m^{2})^{5}}\Big[(p^{2}-m^{2})^{2}\eta^{\mu\nu}\varepsilon^{\alpha\beta}+\nonumber\\&&2(p^{2}-m^{2})\big[\eta^{\mu\nu}p^{\beta}p_{\lambda}\varepsilon^{\alpha\lambda}-2p^{\mu}p^{\nu}\varepsilon^{\alpha\beta}+m^{2}\eta^{\mu\nu}\varepsilon^{\alpha\beta}\big]-
8m^{2}p^{\mu}p^{\nu}\varepsilon^{\alpha\beta}\Big].
\eea
Also, by symmetry reasons we can omit all terms which are
odd in the internal momentum $p$.
Notice that by momentum power counting the above integrals in (\ref{i1}) { involve} only convergent contributions.
Hence, { there is no need to adopt any regularization scheme in order to calculate these integrals. 
They can be readily performed} by using the following formulas:
\bea\label{c1}
&&\int \frac{d^{2}p}{(2\pi)^{2}}\frac{p^{\mu}p^{\nu}}{(p^{2}-m^{2})^{n}}=\frac{(-1)^{n-1}i}{4\pi}\frac{\eta^{\mu\nu}}{2\Gamma(n)}\frac{\Gamma(n-2)}{(m^{2})^{n-2}},\nonumber\\
&&\int \frac{d^{2}p}{(2\pi)^{2}}\frac{p^{\mu}p^{\nu}p^{\rho}p^{\sigma}}{(p^{2}-m^{2})^{n}}=\frac{(-1)^{n}i}{4\pi}\frac{G^{\mu\nu\rho\sigma}}{4\Gamma(n)}\frac{\Gamma(n-3)}{(m^{2})^{n-3}}
\eea
where $G^{\mu\nu\rho\sigma}=\eta^{\mu\nu}\eta^{\rho\sigma}+\eta^{\mu\rho}\eta^{\nu\sigma}+\eta^{\mu\sigma}\eta^{\nu\rho}$.
In what follows, we despise the terms proportional to $n^ {2} \pa^ {2}$ factor. Therefore, by using the equations (\ref{c1}) into
expression (\ref{i2}), we obtain
\bea\label{m4}
S_{eff}[n,\phi,\chi]= \lambda \int d^{2}x \varepsilon^{\alpha\beta}n_{\alpha}(n\cdot \pa)^{2}\chi(x)\pa_{\beta}\phi(x)
\eea
where $\lambda=\frac{\tilde{e}^{2}g^{\prime}}{12\pi m^{2}}\sim {e}^{2}g^{\prime}$ as long as $m\sim 1/R$ as we anticipated. Thus, now taking the mass $m\to0$ we recover the original massless 2D fermionic sector given in (\ref{fermi4D-R-proj}). Thus, we  have indeed derived a nonzero {\it extended} scalar Myers - Pospelov term at one loop effective action in two dimensions through derivative expansion up to leading order in $g^\prime$-parameter. 

Thus, the full effective Lagrangian allows to propose a new theory in two dimensions whose scalar and fermionic sectors are given by
\bea\label{qed_02}
{\cal L}_{{2D}}= \frac12 \partial_{\mu}\phi\partial^{\mu}\phi + \frac12 \partial_{\mu}\chi\partial^{\mu}\chi + \lambda\,\varepsilon^{\alpha\beta}n_{\alpha}(n\cdot\pa)^{2}\chi\partial_{\beta}\phi+
\bar{\psi}( i \pls - m + g^{\prime}(n\cdot \pa)^{2}\ns)\psi- \tilde{e} \bar\psi (\phi+i\gamma_{5}\chi)\psi.
\eea
Notice that
the fermionic sector of the theory (\ref{qed_02})  has the potential to radiatively induce a finite scalar 2D Myers - Pospelov term of the action in Eq.~(\ref{projCFJ}) --- or Eq.~(\ref{v1}) --- initially constructed through a 4D-2D projection.
\section{Conclusion}\label{s5}
In this paper, we investigate the 4D-2D projection of the Myers - Pospelov QED. One of the most interesting results we found is the reduction of dimension-5 operators to dimension-3 operators. One of the main issues addressed by considering higher dimensional operators concerns to
achieve UV-completion in quantum gravity \cite{horava}, for instance, where both two-dimensional versions of the theory and higher dimensional operators are largely considered. In our present analysis we join together these two possibilities in the same set up by applying dimensional projection. Another
issue we considered was the possibility of these Lorentz-violating operators to develop new defect structures in the 2D scalar sector that can be both projected out from the 4D theory and radiatively induced in two dimensions.  

{\bf Acknowledgements.} We would like to thank CNPq, CAPES, PNPD/PROCAD-
CAPES.

\end{document}